\documentclass[prd,aps,showpacs,12pt,amssymb,preprintnumbers,amsmath,amsfonts,nofootinbib]{revtex4-1}

\usepackage{latexsym}
\usepackage[latin1]{inputenc}
\usepackage{amsmath}
\newcommand{\ba}{\begin{eqnarray}}
\newcommand{\ea}{\end{eqnarray}}
\newcommand{\be}{\begin{equation}}
\newcommand{\ee}{\end{equation}}
\newcommand{\pa}{\partial}

\newcommand{\nn}{\nonumber}
\newcommand{\q}{\bf q}

\usepackage{hyperref}  

\usepackage{dcolumn}
\usepackage[dvips]{graphicx}
\usepackage{color}
\usepackage{slashed}

\usepackage{ulem}
\definecolor{stcol}{rgb}{1,0,1}

\begin{document}
\date{\today}

\title{Mass corrections to the hard thermal/dense loops} 
\author{Marc Comadran}
\email{marc.comadranc@e-campus.uab.cat}
\affiliation{Instituto de Ciencias del Espacio (ICE, CSIC) \\
C. Can Magrans s.n., 08193 Cerdanyola del Vall\`es, Catalonia, Spain
and \\
 Institut d'Estudis Espacials de Catalunya (IEEC) \\
 C. Gran Capit\`a 2-4, Ed. Nexus, 08034 Barcelona, Spain
}
\author{Cristina Manuel}
\email{cmanuel@ice.csic.es}
\affiliation{Instituto de Ciencias del Espacio (ICE, CSIC) \\
C. Can Magrans s.n., 08193 Cerdanyola del Vall\`es, Catalonia, Spain
and \\
 Institut d'Estudis Espacials de Catalunya (IEEC) \\
 C. Gran Capit\`a 2-4, Ed. Nexus, 08034 Barcelona, Spain
}

\begin{abstract}
We compute corrections to the hard thermal (or dense) loop photon polarization tensor associated to a small mass $m$ of the fermions of an electromagnetic plasma at high temperature $T$ (or chemical potential $\mu$). To this aim we use the on-shell effective field theory, amended with mass
corrections. We also carry out  the computation using transport theory, reaching to the same result. Interemediate steps in the computations reveal the presence of
potential infrared divergencies. We use dimensional regularization, as it is respectful with the gauge symmetry, and
then show that all infrared divergencies cancel in the final result. We compare the mass corrections with both the power and two-loop
corrections, and claim that they are equally important if the mass is soft, that is, of order $e T$ (or $e \mu$), where $e$ is the gauge coupling constant, but are dominat 
if  the mass obeys $ e T < m \ll T $ (or $e \mu < m \ll \mu)$.

\end{abstract}

\maketitle

\section{Introduction}

Relativistic QED and QCD plasmas have attracted the interest of the physics community for their wide range of applications in both cosmological, astrophysical and
also heavy-ion physics. In their weak coupling regime pertubative computations of different physical observables require the resummation of
Feynman diagrams \cite{Pisarski:1988vd,Braaten:1989mz}, the so called hard thermal loops (HTL), to attain a result valid at a certain order in the gauge coupling expansion (see \cite{Ghiglieri:2020dpq} for a review and complete set of basic references).
This makes the studies of relativistic plasmas particularly cumbersome.

For very large values of the temperature $T$ (or of the chemical potential $\mu$), a well-defined hierarchy of energy scales appears in these relativistic plasmas,
that allows for effective field theory descriptions, very similar to those applied for non-thermal physics. In a series of papers \cite{Manuel:2014dza,Manuel:2016wqs,Manuel:2016cit,Carignano:2018gqt,Carignano:2019zsh,Manuel:2021oah},
 the on-shell efffective theory (OSEFT)
has been developed in order to describe the physics of  the hard scales, or scales of order $T$ (or $\mu $), which are 
 on-shell degrees of freedom.  This effective field theory was initially developed to obtain quantum corrections to classical transport equations. Then it was realized that it could be
used  to improve the description of the hard scales of the plasma, and as by- product, also the soft scales of order $eT$ (or $e \mu$), where $e$ is the gauge coupling constant.

The rationale and technical tools used by OSEFT  are the same as  that of other effective field theories, such as high density field theory (HDET) \cite{Hong:1998tn}, or soft collinear effective field theory (SCET) \cite{Bauer:2000ew,Bauer:2000yr}, for example.
After fixing the value of high energy scale, in this case the energy of the (quasi) massless fermion, which is of order $\sim T$ for thermal plasmas, one defines some small fluctuations 
around that scale. Integrating out the high energy modes, one is left with an effective theory for the lower scales or quantum fluctuations. The resulting Lagrangian
is organized as an expansion of operators of increasing dimension over powers of the high energy scale.

In this manuscript we focus our attention to thermal corrections to the HTL photon polarization tensor associated to the fact that fermions on the plasma might not be strictly massless,
but have indeed a small mass $m$ much less than the temperature, $ m \ll T$. This is a realistic assumption, as only in the cosmological epoch before the electroweak phase transition all elementary particles were strictly massless. 
While the power corrections to the HTL photon polarization tensor have been computed with OSEFT 
 in Ref.~\cite{Manuel:2016wqs}, here we will use the OSEFT for the computation of the leading fermion mass
 corrections. 
We also check that the same result is obtained if derived from transport theory. Intermediate steps in the computations reveal the presence of
potential infrared divergencies. A regularization of the momentum integrals is needed. We use dimensional regularization, as it is respectful with the gauge symmetry, and
then show that all infrared divergencies cancel in the final result. We also explain why our results remain valid
in the presence of a chemical potential, even for high values of $\mu$ and $T=0$.

This paper is structured as follows.  In Sec.~\ref{m-OSEFT} we review the OSEFT Lagrangian including small mass corrections. We present the computation of the
Feynman diagrams in OSEFT that provide mass corrections to the photon polarization tensor in Sec.~\ref{diag-sec}. The same result is obtained if computed from transport theory, as shown in Sec.~\ref{m-kinetic}. We then compare our results with both the
power and two-loop  corrections to the HTL in Sec.~\ref{discussion}, and discuss when the mass corrections are the dominant correction to the HTL polarization tensor. We denote with boldface letters 3 dimensional vectors.
Natural units $\hbar =k_B =1$ are used throughout this manuscript.

\section{Small mass corrections to the OSEFT} 
\label{m-OSEFT}

In this Section we derive the OSEFT Lagrangian including mass corrections to the third order in the energy expansion.
Let us briefly discuss how this is achieved.

In the spirit of the OSEFT we split the momentum of the energetic fermion as
\be
q^\mu = p v^\mu + k^\mu \ ,
\ee
where $v^\mu$ is a light-like vector,  $p$ is the high scale, while $k^\mu$ is the so called residual momentum, 
 associated to the quantum fluctuation, and is such  that $ k^\mu \ll p$.
For the antifermion we will write
\be
q^\mu = -p \tilde{v}^\mu + k^\mu \ ,
\ee
where $ \tilde{v}^\mu$ is also a light-like vector. We will impose that 
\be
 u^\mu = \frac{ v^\mu + \tilde{v}^\mu  }{2}   ,
\ee
where $u^\mu$ is a  frame vector, such that $u^2 =1$, thus, it is time-like.

The OSEFT Lagrangian including small mass corrections has been derived in Ref.\cite{Manuel:2021oah}, and in an arbitrary frame
it reads
\begin{eqnarray}
\mathcal{L}_{p, v}& = &
\bar \chi_{v}(x) \left(i\, v\cdot D\,
+i \slashed{D}_{\perp}  \frac{1}{2 p+ i \tilde{v}\cdot D   }i \slashed{D}_{\perp}  -m^2 
 \frac{1}{2 p + i \tilde{v}\cdot D   }
\right)  \frac{ \slashed{\tilde v} }{2}  \chi_{v}(x) 
\nonumber \\
&-& \bar \chi_{v}(x) \left(  m \Big [ \frac{1}{2 p + i \tilde{v}\cdot D   },   i \slashed{D}_{\perp}   \Big  ]
\right)\frac{ \slashed{\tilde v} }{2}  \chi_{v}(x) 
\ , \label{Leff-OSEFT-chiral}
\end{eqnarray}
for the particle field $\chi_v$ , where   $\slashed{D}_{\perp} = P^{\mu \nu}_{\perp} \gamma_\mu D_\nu$, and the transverse  projector is defined as
$P^{\mu \nu}_{\perp} = g^{\mu \nu} - \frac 12 \left( v^\mu {\tilde v}^\nu +v^\nu {\tilde v}^\mu\right)$.
For antiparticles, the Lagrangian can be obtained after replacing $p \leftrightarrow -p$ and $v^\mu \leftrightarrow \tilde{v}^\mu$.
The Lagrangian has the same structure of that of SCET amended with small mass corrections \cite{hep-ph/0505030,Leibovich:2003jd}.
Note that OSEFT and SCET are different theories, as the power counting is not the same 
(see \cite{Carignano:2018gqt} for a discussion on this
point).

In writing the above Lagrangian, one assumes that the covariant derivatives, defined as $i D_\mu =  i \partial_\mu +  eA_\mu$, are {\it soft}, meaning that they are
much less that the high energy scale, which here it is $p$. Equally, one assumes that the mass is such that $m \ll p$. 
The Lagrangian can then be now expanded using that $p$ is the hard scale of the problem. 

For applications of plasma physics in thermal field theory, it is convenient to use the frame at rest with the plasma, thus $u^\mu= (1, {\bf 0})$.
Then one can replace $\frac{\slashed{\tilde v} }{2}$ by $  \slashed{ u} = \gamma_0$ (recall that  $\slashed{ v}  \chi_{v}= 0$). Expanding the Lagrangian
on the high energy scale, one gets easily the first two terms  in the energy expansion, which respect chirality
\begin{eqnarray}
\mathcal{L}_{p,v}^{(0)} &=&\bar \chi_{v} 
\left( i \, v\cdot D  \right)
\label{Lan-0}
\gamma^0 \chi_{v}  \ , \\
\mathcal{L}_{p,v}^{(1)} &= &-\frac {1}{2 p} \bar \chi_{v} \left( D_{\perp}^{2} + m^2 - \frac e2 \sigma^{\mu \nu}_\perp F_{\mu \nu} \right)   \gamma^0 \chi_{v} \,
\label{Lan-1} \ .
\end{eqnarray}

It is  convenient to introduce local field redefinitions to eliminate the temporal derivative appearing at second order, as in Ref.~\cite{Manuel:2016wqs}. These simplify the computations at higher orders.
Thus, after the field redefinition
\begin{equation}
\label{LFR}
\chi_{v}\rightarrow\chi_{v}^{\prime}=\left(1 +\frac{   \slashed{D}_{\perp}^{2}+ m^2}{8p ^{2}}\right)\chi_{v}\ , 
\end{equation}
the Lagrangian at second order reads
\be
\label{Lan-2}
\mathcal{L}_{p,v}^{(2)} = \bar \chi'_{v} \frac{1}{8p^{2}}  \Big(\left[ \slashed{D}_{\perp}\,,\,\left[i\tilde{v}\cdot D\,,\,\slashed{D}_{\perp}\right]\right] -
\left\lbrace    \slashed{D}_\perp^2+ m^2 ,\,i v\cdot D-i\tilde{v}\cdot D\right\rbrace  + 2 i e m \, \tilde {v}^\mu F_{\mu \alpha} \gamma^\alpha_\perp  
\Big)
 \gamma^0\chi'_{v}\ . 
\ee
Note that  the term linear in the mass describes  a breaking of chirality.

We will also need the Lagrangian up to third order. To eliminate temporal derivatives at that order, we perform the local field redefinition
\ba
\chi_v  \rightarrow \chi_v'' &=& \left( 1-\dfrac{i}{8 p^3}\slashed{D}_\perp[\tilde{v}\cdot D,\slashed{D}_\perp]+\dfrac{i}{16 p^3}(\slashed{D}_\perp^2+m^2)\left( v\cdot D-\tilde{v}\cdot D\right)-\dfrac{i}{16 p^3}(\slashed{D}_\perp^2+m^2)\tilde{v}\cdot D \right.
\nonumber  
\\
&+& \left. \frac{m}{8p^3} \left[ i\tilde{v}\cdot D\,,\,i \slashed{D}_{\perp} \right]
\right)\chi'_v \ ,
\ea
so that the final Lagrangian reads
\begin{eqnarray}
\label{Lan-3}
\mathcal{L}^{(3)}_{p,v} &=&\dfrac{1}{8p^3} \bar \chi_v^{''} \Bigg( (\slashed{D}_\perp^2+m^2)^2 +[i\tilde{v}\cdot D,\slashed{D}_\perp]^2-(iv\cdot D-i\tilde{v}\cdot D)(\slashed{D}_\perp^2+m^2)(iv\cdot D-i\tilde{v}\cdot D) 
\nonumber \\
&+&   (iv\cdot D-i\tilde{v}\cdot D)\slashed{D}_\perp[i\tilde{v}\cdot D,\slashed{D}_\perp]-[i\tilde{v}\cdot D,\slashed{D}_\perp]\slashed{D}_\perp (iv\cdot D-i\tilde{v}\cdot D)
\nonumber
 \\
&+&  m \left\{ i v\cdot D-i\tilde{v}\cdot D,\left[ i\tilde{v}\cdot D\,,\,i \slashed{D}_{\perp} \right] \right\} 
  \Bigg )\gamma^0\chi_v^{''} .
\end{eqnarray}

Please note that in the limit $m=0$ we recover the same Lagrangians derived in Ref.~\cite{Manuel:2016wqs}.
The pieces which are quadratic in the mass can be recovered from those of   Ref.~\cite{Manuel:2016wqs} simply by replacing
$\slashed{D}_\perp^2 \rightarrow  \slashed{D}_\perp^2 + m^2$. The linear terms in $m$, originating from the expansion of the  last term in 
Eq.~(\ref{Leff-OSEFT-chiral}),  describe  the breaking of the chiral symmetry induced by the fermion mass.

We present here how the OSEFT fermion propagators are modified in the presence of a small mass. The particle/antiparticle projectors in the frame at rest with the plasma are defined as $P_v = \frac 12 \slashed{v}  \gamma_0$ and $P_{\tilde v} = \frac 12 \slashed{\tilde v}  \gamma_0$, respectively. 
We  introduce chirality projectors
\be
P_\chi = \frac{1 + \chi \gamma_5}{2} \ , \qquad \chi = \pm \ . 
\ee

The propagators for a fermion of chirality $\chi$ in the Keldysh formulation of the real time formalism of thermal field theory Ref.~\cite{Chou:1984es} read
\begin{eqnarray}
\label{RAF-propa}
S^{R/A}_\chi(k) & = &  \frac{ P_\chi P_v \gamma_0   }{k_0  \pm i \epsilon - f({\bf k},m)} ,\\
S^{S}_\chi(k)  & = & P_\chi P_v \gamma_0 \left( - 2\pi i \delta( k_0 - f({\bf k},m)) \left( 1 - 2n_F^\chi(p +k_0)  \right) \right)  \, ,
\end{eqnarray}
where $n_F (x)= 1/(\exp{|x|/T} +1)$ is the Fermi-Dirac equilibrium distribution function.
The function $f({\bf k},m)$ determines the dispersion law, and it is 
 expanded also, we denote as $f^{(n)}({\bf k},m)$ the $n$ order term in the $1/p$ expansion. At lowest order
\be
f^{(0)}({\bf k},m) =   k_\parallel \ ,
\ee
and we have defined $ k_\parallel = {\bf k} \cdot {\bf v}$, while 
\be
f^{(1)}({\bf k},m) = k_\parallel + \frac{{\bf k}_\perp^2+m^2}{2 p} \ , \qquad
 f^{(2)}({\bf k},m) = k_\parallel + \frac{{\bf k}_\perp^2+m^2}{2 p} - \frac{k_\parallel ({\bf k}_\perp^2+m^2)}{2 p^2 } \ ,
 \label{displaw-2}
  \ee
as follows from Eqs.~(\ref{Lan-1}) and (\ref{Lan-2}), respectively. 
The propagators for the antiparticle quantum fluctuations can be
also be easily deduced. They  read
\begin{eqnarray}
\label{RAFanti-propa}
{\widetilde S}^{R/A}_\chi(k) & = &  \frac{ P_\chi P_{\tilde v} \gamma_0   }{k_0  \pm i \epsilon - {\tilde f}({\bf k},m)} ,\\
{\widetilde S}^{S}_\chi(k)  & = & - P_\chi P_{\tilde v} \gamma_0 \left( - 2\pi i \delta( k_0 - {\tilde f}({\bf k},m)) \left( 1 - 2n_F(-p +k_0)  \right) \right)  \, ,
\end{eqnarray}
where the function ${\tilde f}({\bf k},m)$ can be obtained from $f({\bf k},m)$, with the replacements ${\bf v} \rightarrow - {\bf v}$ and $p  \rightarrow -p$.
Note the extra minus sign in the symmetric antiparticle propagator, absent in its particle counterpart.

In summary, the OSEFT fermion propagators in this case can be deduced from those of the massless case simply by replacing ${\bf k}_\perp^2 \rightarrow
{\bf k}_\perp^2+m^2$ in the function  that determines the dispersion relation at every order in the energy expansion.

Note that, for convenience, we keep the propagators above unexpanded in this Section, as done in
Ref.~\cite{Manuel:2016wqs}, but in the explicit computation of the different diagrams they are to be expanded in a $1/p$ series.

\section{Diagrammatic computation of the  mass correction to the retarded photon polarization tensor}
\label{diag-sec}

\begin{figure}
	\centering
	\DeclareGraphicsExtensions{.pdf}
	\includegraphics[scale = 0.75]{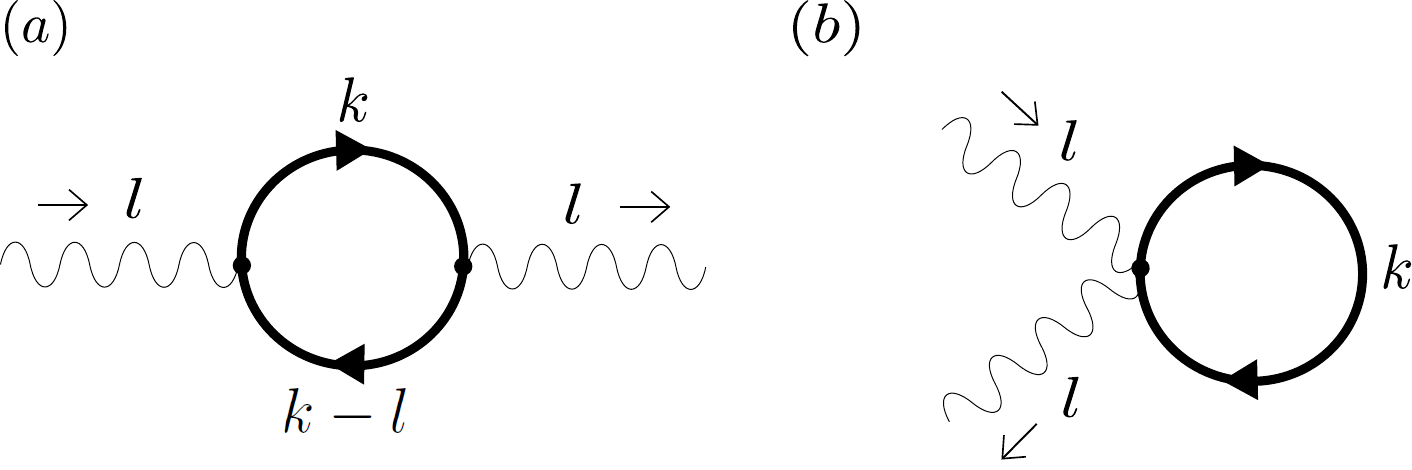}
	\caption{$(a)$ Bubble diagram $(b)$ Tadpole diagram.}
	\label{Diagrams}
\end{figure}

In this Section we compute the mass corrections to the retarded photon polarization tensor computed in OSEFT.
Recall that there are two possible different topological diagrams that contribute to the computation, the bubble and the tadpole diagrams, see Fig.~\ref{Diagrams}.
The tadpole diagrams, absent in QED, take into account particle-photon interactions mediated by an off-shell antiparticle
(and viceversa for antiparticle-photon interactions), and are needed to respect the Ward identity obeyed by the polarization tensor, as we will
explicitly check in this manuscript.

In the Keldysh representation the  particle contribution to the bubble diagram  of the retarded polarization tensor has the structure 
\footnote[1]{We have changed the sign conventions of the definition of the polarization tensor as with respect to those used in Ref.~\cite{Manuel:2016wqs}.}
\begin{equation}
\Pi_{b,\chi}^{\mu\nu}(l)=\dfrac{i}{2}\sum_{p,{\bf v}} \int\dfrac{d^4k}{(2\pi)^4}\Big\lbrace \text{Tr}\big[V^\mu S^\chi_S(k-l)V^\nu S^\chi_R(k)\big]
+\text{Tr}\big[V^\mu S^\chi_A(k-l)V^\nu S^\chi_S(k)\big]\Big\rbrace \ ,
\end{equation}
while the particle contribution to the tadpole diagram can be expressed as
\begin{equation}
\Pi_{t,\chi}^{\mu\nu}(l)=- \dfrac{i}{2}\sum_{p,{\bf v}} \int\dfrac{d^4k}{(2\pi)^4}\text{Tr}\left[ W^{\mu\nu} S^\chi_S(k)\right] \ ,
\end{equation}
where the momentum dependence of the vertex functions $V^\mu$ and $W^{\mu \nu}$ are understood.
Similar expressions can be written for the antiparticle contributions to the polarization tensor.

Using the explicit expressions of the fermion propagators, one can carry out the integral in $k_0$ to arrive
to the general expressions
\begin{equation}
\Pi_{b,\chi}^{\mu\nu}(l)=- \sum_{p,{\bf v}}\int\dfrac{d^3 {\bf k}}{(2\pi)^3} \text{Tr}\big[V^\mu P_\chi P_v\gamma^0 V^\nu  P_v\gamma^0\big]\dfrac{n_F(p+f({\bf{k} -\boldsymbol{l}},m))-n_F(p+f({\bf k},m))}{l_0+ i 0^+ +f({\bf k- \boldsymbol{l}},m)-f({\bf k},m)} \ ,
\end{equation}
and
\begin{equation}
\Pi_{t,\chi}^{\mu\nu}(l)=  - \frac12 \sum_{p,{\bf v}}\int\dfrac{d^3 {\bf k}}{(2\pi)^3}\text{Tr}\left[  W^{\mu\nu} P_\chi P_v\gamma^0\right]\Bigg( 1 - 2 n_F(p+f({\bf k},m)) \Bigg) \ ,
\end{equation}
for the bubble and tadpole diagrams,
respectively. 

The Feynman rules needed for the computation of the photon polarization tensor were given in Ref.~\cite{Manuel:2016wqs} (see Tables I and II of that reference). In the presence of
a mass in the OSEFT Lagrangian, new vertices appear proportional to the mass squared, which are given by
\begin{eqnarray}
V_{(2),m^2}^{\mu} & = & -\frac{e m^2}{2p^{2}}\gamma_{0} \, \delta^{\mu i}v^{i} \ , 
  \\
W^{\mu\nu}_{(3),m^2} & = &  -\dfrac{m^2e^2\gamma^0}{2p^3}\left[ P_{\perp}^{\mu\nu}+\dfrac{(v^\mu -\tilde{v}^\mu)(  v^\nu -\tilde{v}^\nu )}{2}\right] \ .
\end{eqnarray}

There are also new vertices proportional to the mass, which imply a change in the fermion chirality. At the order we will compute the mass corrections, $n=3$ in the
energy expansion, these will not be needed, although they would be required at fourth order in the energy expansion. Note that at least two of these vertices would
be needed in a computation of the photon polarization tensor to preserve the fermion chirality inside the loop.

We now evaluate the polarization tensor at different orders, noting that we either consider the energy expansion in the vertex functions, or in the fermion propagators, which can be used at  the desired order of accuracy.

The first non-vanishing contribution to the photon polarization tensor occurs at $n=1$, but it does not carry any mass dependence.
This was computed in  Ref.~\cite{Manuel:2016wqs}, and it reproduces the HTL contribution. Let us recall the main results here.
 Adding the bubble and the tadpole diagrams at order $n=1$ gives
\begin{equation}
\Pi^{\mu\nu} (l)= - e^2 \sum_{\chi= \pm}  \sum_{p,{\bf v}}\int\dfrac{d^3 {\bf k}}{(2\pi)^3}\dfrac{dn_F}{dp}\left( \dfrac{P^{\mu\nu}_\perp}{2}+v^\mu v^\nu -l_0\dfrac{v^\mu v^\nu}{v\cdot l}\right) .
\end{equation}
where the retarded prescription $l_0 \rightarrow l_0 +  i 0^+$ is understood.

It is now important to return to the original momentum variable $q^\mu$. Using the identity \cite{Manuel:2016wqs}
\be
\sum_{p,{\bf v}}\int\dfrac{d^3 {\bf k}}{(2\pi)^3} \equiv \int\dfrac{d^3 {\bf q}}{(2\pi)^3}
\ee 
and the relations \cite{Manuel:2016wqs}
\begin{eqnarray}
\label{p-back}
p &=& q - k_{\parallel,{\bf \hat q}} + \frac{{\bf k}_{\perp,{\bf \hat q}}^2 }{2 q} +{\cal O}\left( \frac{1}{q^2}\right)  \ ,\\
\label{v-back}
{\bf v} & = & {  \bf{\hat q}} -  \frac{{\bf k}_{\perp,{\bf \hat q}} }{q}  - \frac{  {\bf \hat q} {\bf k}_{\perp,{\bf \hat q}}^2 + 2  k_{\parallel,{\bf \hat q}} {\bf k}_{\perp,{\bf \hat q}}}{2 q^2}+{\cal O}\left( \frac{1}{q^3}\right) 
\ , \\
\label{n-back}
n_F(p) &= & n_F(q) +  \frac{d n_f}{dq} \left( -  k_{\parallel,{\bf \hat q}}  + \frac{{\bf k}_{\perp,{\bf \hat q}}^2 }{2 q} \right) + \frac 12 \frac{d^2 n_F}{dq^2}   k_{\parallel, {\bf \hat q}}^2 +
{\cal O}\left( \frac{1}{q^3}\right)  \ ,
\end{eqnarray} 
where now the symbols ${\bf k}_{\parallel, {\bf \hat q}}$ and ${\bf k}_{\perp,{\bf \hat q}}$ denote the components 
of ${\bf k}$ parallel and perpendicular to $\hat{{\q}}\equiv {\q}/q$ with $q=\vert \bf q\vert$, respectively.
We also define the vectors 
\be
v_{\hat{\q}}^\mu \equiv (1 ,{\hat{\bf q}}) \ , \qquad {\tilde v}_{\hat{\q}}^\mu \equiv (1 ,-{\hat{\bf q}}) \ .
\ee

After adding both the particle and antiparticle contributions to the photon polarization tensor one arrives to the well-known HTL expression
\begin{equation}\label{4.20}
\Pi^{\mu\nu}_{{\rm htl}}(l) = - 4   e^2    \int\dfrac{d^3 {\bf q}}{(2\pi)^3} \dfrac{dn_F}{dq}
\left( \delta^\mu_0 \delta^\nu_0 -l_{0}  \dfrac{v_{\hat {\q}}^\mu v_{\hat {\q}}^\nu}{v_{\hat{\q}}\cdot l} \right) \ .
\end{equation}

At second order in the energy expansion, and in the absence of chiral misbalance, the Bose-Einstein statistics and the crossing symmetry demands that
the polarization tensor $\Pi^{\mu \nu} (l)$  be symmetric under the simultaneous exchange of $\mu \leftrightarrow \nu$ and $l \leftrightarrow -l$ \cite{Nieves:1988qz}.
These symmetries explain the absence of linear terms in the photon momenta in the polarization tensor, which ultimately explain why there are not $n=2$ corrections
in the polarization tensor in OSEFT. This was explicitly checked in  Refs.~\cite{Manuel:2016wqs,Carignano:2017ovz}. This reasoning applies actually to all the
even orders of the energy expansion, $n=4,6, 8, \ldots$. We do not expect thus mass corrections at even orders, neither, and we actually have checked that there are none
at $n=2$.

The first mass corrections to the photon polarization tensor occur at third order in the energy expansion, the same as the power corrections to the HTL
computed in Refs.~\cite{Manuel:2016wqs,Carignano:2017ovz}.

 The mass dependent terms that arise in the bubble diagram are
\begin{eqnarray}
\Pi_{b}^{\mu\nu}(l)
& = & -m^2e^2  \sum_{\chi= \pm}\sum_{p,{\bf v}}\int\dfrac{d^3 {\bf k}}{(2\pi)^3}\left\lbrace 
-\dfrac{l_\parallel}{2p}\left[\left(  \dfrac{d^2n_F}{dp^2}-\dfrac{1}{p}\dfrac{dn_F}{dp}\right) \dfrac{v^\mu v^\nu}{v\cdot l}-\dfrac{l_\parallel}{p} \dfrac{dn_F}{dp}\dfrac{v^\mu v^\nu}{(v\cdot l)^2} \right. \right.
\nonumber \\
&- & 
\left. \left.
\dfrac{1}{2 p}\dfrac{dn_F}{dp}\dfrac{v^\mu (v^\nu - {\tilde v}^\nu )+ v^\nu (v^\mu - {\tilde v}^\mu)  }{v\cdot l}\right]\right\rbrace   \ , 
\end{eqnarray}
while in the tadpole one gets
\begin{equation}
\Pi_{t}^{\mu\nu}(l)
= - m^2e^2  \sum_{\chi= \pm}\sum_{p,{\bf v}} \int\dfrac{d^3 {\bf k}}{(2\pi)^3}\left\lbrace
\dfrac{n_F}{2p^3}\left[ P_{\perp}^{\mu\nu}+\dfrac{(v^\mu  - \tilde{v}^\mu )(v^\nu - \tilde{v}^\nu)}{2}\right]- \dfrac{P^{\mu\nu}_\perp}{2p^2}\dfrac{dn_F}{dp}
\right\rbrace.
\end{equation}
We add the  two pieces, and go back  to the full momentum variables.
The final result, after adding also the antiparticle contributions, yields 
\begin{equation}
\begin{gathered}
\label{bubl+tapd}
\Pi_{{\rm m}}^{\mu\nu}(l)= - 4 m^2e^2 \int\dfrac{d^3 {\bf q}}{(2\pi)^3}\bigg\lbrace
\dfrac{n_F}{2q^3}\left[ P_{\perp,\hat{\q}}^{\mu\nu}+\dfrac{   (v_{\hat{\q}}^\mu  - \tilde{v}_{\hat{\q}}^\mu )(v_{\hat{\q}}^\nu - \tilde{v}_{\hat{\q}}^\nu)  
}{2}\right]
\\
+\dfrac{1}{q^2}\dfrac{dn_F}{dq}\left[- \dfrac{P^{\mu\nu}_{\perp,\hat{\q}}}{2} +l_{\parallel,\hat{\q}}\left( \dfrac{v^\mu_{\hat{\q}} v^\nu_{\hat{\q}}}{v_{\hat{\q}}\cdot l}+\dfrac{l_{\parallel,\hat{\q}}}{2} \dfrac{v^\mu_{\hat{\q}} v^\nu_{\hat{\q}}}{(v_{\hat{\q}}\cdot l)^2}+\dfrac{1}{4}\dfrac{v^\mu_{\hat{\q}} (v^\nu_{\hat{\q}} -\tilde{v}^\nu_{\hat{\q}})
+v^\nu_{\hat{\q}} (v^\mu_{\hat{\q}} -\tilde{v}^\mu_{\hat{\q}})
}{v_{\hat{\q}}\cdot l}\right) \right] \bigg\rbrace \ .
\end{gathered}
\end{equation}

We note that the first two terms of the second line of Eq.~({\ref{bubl+tapd}) can be written as the HTL contribution, but with a coefficient proportional
to $e^2 m^2$ rather than the Debye mass squared $m^2_D = e^2 T^2/3$. Note also that  in the tadpole diagram the pieces that are proportional to  $n_F(q)/q^3$ are
in principle infrared divergent. These terms have to be evaluated using a regularization. We use dimensional regularization (DR), by assuming that
the system is in $d=3 + 2 \epsilon$ dimensions. In this case the momentum integrals become
\be
\int \frac{d^d q}{(2\pi)^d}  \rightarrow 
\frac{ 4 }{(4\pi)^{2 + \epsilon} \Gamma( 1+ \epsilon)} \,\int_0^\infty dq \, q^{2+ 2 \epsilon} \int_{-1}^1 d\cos\theta \left( 1 +  \epsilon \ln{(\sin^2 \theta)} \right) \ ,
\ee
where $\theta$ parametrises an angle with respect to an external vector, and $\Gamma(z)$ stands for the Gamma function.
Furthermore, in $d$ dimensions one has to change the coupling constant as $e^2 \rightarrow e^2 \nu^{3-d}$,
where $\nu$ is a renormalization scale.

The relevant infrared radial integral is
\begin{align}
 \nu^{- 2 \epsilon} \int_0^\infty dq q^{-1+2\epsilon} \,n_F (q)  & = 
\frac{1}{4\epsilon} + \frac{1}{2}\ln \left(\frac{\pi Te^{-\gamma_E}}{2 \nu}\right)  + {\cal{O}}(\epsilon) \  ,
\label{DR-fer}
\end{align}
where $\gamma_E$ is Euler's constant.
However, when carrying out the angular integrals in $d= 3 + 2 \epsilon$ dimensions, the pole term and logarithm exactly cancel, as the angular integral turns out to
be proportional to $\epsilon$ (that is, it would cancel if $d=3$). If $d \Omega_d$ is the solid angle element in $d$ dimensions, and $S_d= 2 \pi^d/ \Gamma(d/2)$ is the area of a $d$-dimensional unit sphere, one can check
\begin{equation}
S^{-1}_{3 + 2 \epsilon} \int d \Omega_{3 + 2 \epsilon} \left( - \delta^{i j} + 3 \hat{\q}^i\hat{\q}^j \right) = - \frac{2}{3} \epsilon  + {\cal{O}}(\epsilon^2) \ .
\end{equation} 
Thus, combing the two results one gets 
\begin{equation}
\label{IR-bubble}
- 4 m^2e^2  \nu^{- 2 \epsilon} \int\dfrac{d^d q}{(2\pi)^d}
\dfrac{n_F}{2q^3}\left[ P_{\perp,\hat{\q}}^{\mu\nu}+\dfrac{(v_{\hat{\q}}^\mu  - \tilde{v}_{\hat{\q}}^\mu )(v_{\hat{\q}}^\nu - \tilde{v}_{\hat{\q}}^\nu) } 
{2}\right] = \dfrac{m^2e^2}{6\pi^2}\delta^{ij}  + {\cal{O}}(\epsilon)\ ,
\end{equation}
and there is no infrared divergence, but only a finite term. This finite term is ultimately needed
to preserve the Ward identity obeyed by the polarization tensor, as can be checked after computing
\ba
\nn
l_\mu  \Pi_{{\rm m}}^{\mu\nu}(l) &=&  4 m^2e^2  \int\dfrac{d^3 {\bf q}}{(2\pi)^3}\dfrac{1}{q^2}\dfrac{dn_F}{dq} \left[
\dfrac{l^2_{\parallel,\hat{\q}}}{2} \dfrac{ v^\nu_{\hat{\q}}}{v_{\hat{\q}}\cdot l}-\dfrac{2 l^2_{\parallel,\hat{\q}}}{4}\dfrac{v^\nu_{\hat{\q}}} {v_{\hat{\q}}\cdot l} 
+\frac {l_{\parallel,\hat{\q}}}{4}   (v_{\hat{\q}}^\nu - \tilde{v}_{\hat{\q}}^\nu ) \right] 
\\
&-&\dfrac{m^2e^2}{6\pi^2} l^j \delta^{\nu j}  =0 \ .
\label{WId}
\ea

Note that if we had used a cutoff regularization of the integrals, the IR divergent terms would also vanish, but the above integral would not
yield the finite contribution, the last term of Eq.~(\ref{WId}),  needed to respect the gauge invariance of the computation.

We define the longitudinal and transverse parts of the photon polarization tensor in $d$ dimensions by
\be
\Pi^L (l_0,\boldsymbol{l})\equiv \Pi^{00}(l_0,\boldsymbol{l}) \ , \qquad \Pi^T(l_0,\boldsymbol{l}) \equiv \frac{1}{d-1} \left(\delta^{ij} - \frac{l^il^j}{\boldsymbol{l}^2} \right) \Pi^{ij}(l_0,\boldsymbol{l}) \ .
\ee

We then find the following mass corrections to the longitudinal and transverse parts of the polarization tensor
\ba
\label{final-L}
\Pi_{\rm m}^{\rm L} (l_0,\boldsymbol{l})& = &   \frac{e^2 m^2}{2 \pi^2} \frac{ \boldsymbol{l}^2 }{ l^2_0 - \boldsymbol{l}^2 } \ , 
 \\
 \label{final-T}
\Pi_{\rm m}^{\rm T} (l_0,\boldsymbol{l}) & =&     \frac{e^2 m^2}{2 \pi^2} \frac{ l_0 }{2 \vert \boldsymbol{l} \vert} \ln \left( { \frac{l_0 +\vert \boldsymbol{l} \vert }{l_0-\vert \boldsymbol{l} \vert}   }\right)  \ . 
\ea

Let us finally stress that Eqs.~(\ref{final-L})-(\ref{final-T}) remain also valid in the presence of a finite chemical potential $\mu$.
In the presence of a chemical potential the particle and antiparticle contributions differ, but the final result can be recovered from 
Eq.~(\ref{bubl+tapd}), simply by replacing in Eq.~(\ref{bubl+tapd})
\be
n_F(q) \rightarrow \frac 12 \left[ n_F(q -\mu) + n_F(q + \mu)   \right]  \ . 
\ee

After an explicit evaluation of the corresponding integrals, one reaches to the same mass corrections to the polarization tensor which
are valid at high temperature. In particular, our results still hold if we take $T= 0$ and keep the chemical potential
$\mu$ as the high scale of the problem.

\section{Computation of the photon polarization tensor from kinetic theory}
\label{m-kinetic}

We compute in this Section the mass corrections to the photon polarization tensor as computed from kinetic theory.
We use the transport approach derived from OSEFT, and focus on the vectorial component of the Wigner
function.
From Ref.~\cite{Manuel:2021oah}, the transport equation associated to a fermion with chirality $\chi$
up to second order in the energy expansion  reads

\begin{align}
 \left[  v^\mu_\chi- \frac{e}{2 E^2_q}S_\chi^{\mu \nu}F_{\nu \rho}  (X) \left(2 u^\rho - v^\rho_\chi \right) \right] 
\left(\pa_{\mu}^X  -e  F_{\mu \rho} (X) \pa^\rho_{q} \right)
G^\chi(X,q)  &  = 0 \ , \label{CCTEq-1} 
 \end{align}
where $v^\mu_\chi = q^\mu/E_q$,  and we take the frame vector that defines the system as $u^\mu=(1,{\bf 0})$. Furthermore
\be
G^\chi(X,q) =  2\pi \delta(Q^\chi_m) n^\chi(X,q) \ ,
\ee 
where $n^\chi(X,q)$ is the distribution function, and  the delta gives the on-shell constraint, $Q^\chi_m$ being
a function of the momentum and the mass.
 The  particle contribution to the electromagnetic current is expressed, at $n=2$ order as
\begin{align}
j^\mu(X) & =  e \sum_{\chi=\pm} \int \frac{d^4q}{(2 \pi)^4}   \left[ v_\chi^\mu - \frac{S_\chi^{\mu\nu} \Delta_\nu}{E_q}-\frac{e}{2} \frac{S_\chi^{\mu\nu}}{E_q^2} F_{\nu\rho}(X) (2u^\rho-v_\chi^\rho) \right] 2G^\chi(X,q)  + {\cal O} \left(\frac {1}{E_q^3}  \right)  \ .
\end{align}

We ignore in this manuscript the possible effect of the spin coherence function discussed in Ref.~\cite{Manuel:2021oah}, which represent coherent quantum states of mixed chiralities.
We also will ignore the terms in the transport equation, on-shell constraint,   and  in the vector current proportional to the spin tensor $S_\chi^{\mu \nu}$, as they are irrelevant if the chiral chemical potential is zero, as the contribution of the two fermion chiralities makes these pieces to cancel in the macroscopic current. Those terms are relevant, though, to derive
the chiral magnetic effect, which is not our goal here (see Appendix of Ref. \cite{Carignano:2018gqt} for that derivation).
We thus write the on-shell constraint  to the considered order of accuracy
\be
q_0 = E_q = q + \frac{m^2}{2 q}  \ .
\ee

Where $q=\vert \bf q \vert$. We now assume to be close to thermal equilibrium, such that
\be
G^\chi = G_{(0)}^\chi + \delta G^\chi + \cdots
\ee
where $G_{(0)}^\chi$  is the Wigner function in thermal equilibrium. Using the transport equation, we find

\begin{align}
   v^\chi \cdot \pa_X   \delta G^\chi =  e  v^\chi_\mu F^{\mu \nu} \,\frac{\pa G^\chi_{(0)}}{\pa q^\nu}
\ ,
 \end{align}
and after computing
\be
\delta j^\mu(X)  =  e \sum_{\chi=\pm} \int \frac{d^4q}{(2 \pi)^4}  v_\chi^\mu  \,\delta G^\chi(X,q) \ ,
 \ee
one derives the polarization tensor as
\be
\Pi^{\mu \nu} = \frac{\delta j^\mu}{\delta A_\nu} \ .
\ee

It is not difficult to find  the particle contribution to the polarization tensor, which reads
\be
\Pi^{\mu \nu} (l) = e^2 \sum_{\chi=\pm} \int  \frac {d^3 {\bf q}}{(2 \pi)^3 }\left ( g^{\mu \nu} - \frac{l^\mu v_m^\nu + v_m^\mu l^\nu}{ l \cdot v_ m} + L^2 \frac{v_m^\mu v_m^\nu}{(l \cdot v_m)^2} \right)\frac{n_F(q_0=E_q)}{E_q} \ ,
\ee
where $L^2 = l_0^2 - \boldsymbol{l}^2$, and
\be
v^\mu_m = v^\mu_{\hat{\q}} - \delta^{\mu i} v^i _{\hat{\q}} \frac{m^2}{2q^2} \ ,
\ee
for the particles. A similar expression holds for the antiparticles.

We compute all the pieces up to ${\cal O}(m^2)$, by noting that
\ba
\frac{1}{E_q} &=& \frac 1q - \frac{m^2}{2 q^3} + \cdots\\
\frac{1}{l \cdot v_m} &=& \frac{1}{l \cdot v_{\hat{\q}}} - \frac{{\boldsymbol{l}}\cdot {\bf v}_{\hat{\q}} }{(l \cdot v_{\hat{\q}})^2 }  \frac{m^2}{ 2 q^2} + \cdots \\
n_F(E_q) &=& n_F(q) + \frac{m^2}{2 q } \frac{ d n_F}{dq}+ \cdots 
\ea

We thus find that the polarization tensor can be written as 
\be
\Pi^{\mu \nu} (l) = \Pi_{\rm htl}^{\mu \nu} (l) + \Pi_{\rm m}^{\mu \nu} (l)  \ .
\ee
The HTL part, as arising from particles and antiparticles of the two possible chiralities, reads
\be
\Pi_{\rm htl}^{\mu \nu} (l) = 4 e^2 \int  \frac {d^3 {\q} }{(2 \pi)^3 }\left ( g^{\mu \nu} - 
\frac{l^\mu v_{\hat{\q}}^\nu + v_{\hat{\q}}^\mu l^\nu}{ l \cdot v_{\hat{\q}}} + L^2 \frac{v_{\hat{\q}}^\mu v_{\hat{\q}}^\nu}{(l \cdot v_{\hat{\q}})^2} \right)\frac{n_F (q)}{q}.
\ee
While the leading mass correction is
\ba
\label{m-phpol}
\Pi_{\rm m}^{\mu \nu} (l) &=& 4 e^2 m^2 \int  \frac {d^3 {\q}}{(2 \pi)^3 }\left( g^{\mu \nu} - 
\frac{l^\mu v_{\hat{\q}}^\nu + v_{\hat{\q}}^\mu l^\nu}{ l \cdot v_{\hat{\q}}} + L^2 \frac{v_{\hat{\q}}^\mu v_{\hat{\q}}^\nu}{(l \cdot v_{\hat{\q}})^2} \right)
\left(  \frac{1}{2 q^2} \frac{d n_F(q)}{dq} \right) 
\nonumber
\\
&+& 4 e^2 m^2 \int  \frac {d^3 {\q}}{(2 \pi)^3 } \left( g^{\mu \nu}-  l_0  \frac{l^\mu v_{\hat{\q}}^\nu + v_{\hat{\q}}^\mu l^\nu}{ (l \cdot v_{\hat{\q}})^2 } -L^2  \frac{v_{\hat{\q}}^\mu v_{\hat{\q}}^\nu}{ (l \cdot v_{\hat{\q}})^2 } 
 +2 L^2l_0\frac{v_{\hat{\q}}^\mu v_{\hat{\q}}^\nu}{(l \cdot v_{\hat{\q}})^3} \right.
\nonumber \\
&-& \left.  \frac{\delta^{\mu i} v_{\hat{\q}}^i l^\nu + \delta^{\nu i} v_{\hat{\q}}^i l^\mu}{ l \cdot v_{\hat{\q}}} + L^2  \frac{\delta^{\mu i} v_{\hat{\q}}^i v_{\hat{\q}}^\nu + \delta^{\nu i} v_{\hat{\q}}^i v_{\hat{\q}}^\mu}{(l \cdot v_{\hat{\q}})^2} \right) \left(- \frac{n_F(q)}{2 q^3} \right) \ .
\ea
 One can  check that
\be 
l_\mu \Pi_{\rm m}^{\mu \nu} (l)  = 0 \ ,
\ee
so that the Ward identity is respected for the mass dependent pieces of the polarization tensor at this order.
Note that the first integral of Eq.~(\ref{m-phpol}) 
has the same structure than the HTL contribution, but it is proportional to the fermion mass squared. This contribution
was also found out in the diagrammatic computation of Sec.~\ref{diag-sec}.
The second integral contains IR divergencies, but are of a quite different structure as those appearing in the diagrammatic computation, see
Eq.~(\ref{IR-bubble}). The apparent IR divergencies here are clearly non-local. We evaluate these integrals using
DR.

Note that we only need to evaluate the integral
\begin{align}
I_1 & \equiv  \int_{-1}^1 d\cos\theta \left(1-\cos^2\theta\right)^\epsilon \frac{1}{l_0 - \vert \boldsymbol{l} \vert \cos\theta} \nonumber\\
      &  =  \frac{1}{l} \left\{ \ln \left( \frac{l_0 + \vert \boldsymbol{l} \vert }{l_0 - \vert \boldsymbol{l} \vert }\right)  + \epsilon \left[ \ln (4) {\ln\left( \frac{l_0 + \vert \boldsymbol{l} \vert }{l_0 - \vert \boldsymbol{l} \vert }\right) } + \text{Li}_2 \left(-\frac{2 l}{l_0 - \vert \boldsymbol{l} \vert } \right)- \text{Li}_2\left(\frac{2 l}{l_0 + \vert \boldsymbol{l} \vert }\right) \right] \right\} +  {\cal{O}}(\epsilon^2) \ ,
\end{align}
where $\text{Li}_2$ stands for the Euler polylogarithmic function of order 2.
 All the remaining non-local integrals can de deduced from this one, after simple manipulations.

An explicit computation shows that after angular integration in $d= 3 + 2 \epsilon$ dimensions the IR 
divergencies exactly cancel,  but there are remaining finite pieces, which in this case turn out to be non-local, and that allows one to reproduce the same value of the photon polarization tensor that we found in Sec.~\ref{diag-sec}.

\section{Discussion}
\label{discussion}

We used OSEFT to assess how  a small fermion mass  would affect  the retarded photon polarization tensor at soft scales in a ultrarelativistic electromagnetic plasma. While it could be obvious that such corrections would be of order $m^2/T^2$, the effective field
theory techniques we used allowed us their proper evaluation.

Our results should be  compared to both the power  and two-loop corrections to the HTL  tensor
that have been computed in Refs. \cite{Manuel:2016wqs,Carignano:2017ovz}
and \cite{Carignano:2019ofj}, respectively. More precisely, we will write
 \be 
\label{exp-HTL}
 \Pi_I= \Pi_I^{\rm htl} +\Pi_I^m + \Pi_I^{\rm pow\cdot corr}+\Pi_I^{\rm 2loop} \ , \qquad  I= L, T \ ,
 \ee
where $\Pi_I^m$ were displayed in Eqs.~(\ref{final-L},\ref{final-T}) and 
\ba
 &&\Pi_L^{\rm htl}(l_0,\boldsymbol{l}) = \frac{e^2T^2}{3}\left(1- \frac{l_0}{2\vert \boldsymbol{l} \vert} \ln \left(\frac{l_0+\vert \boldsymbol{l} \vert}{l_0-\vert \boldsymbol{l} \vert}\right)\right) \ ,\nn\\
&& \Pi_L^{\rm pow\cdot corr}(l_0,\boldsymbol{l}) = -\frac{e^2}{4 \pi ^2}
 \Big(\boldsymbol{l}^2  -\frac{l_0^2}{3}\Big)\Big( 1 - \frac{l_0}{2\vert \boldsymbol{l} \vert} \ln\left(\frac{l_0+\vert \boldsymbol{l} \vert}{l_0-\vert \boldsymbol{l} \vert}\right)\Big) \ , \nn\\
&& \Pi_L^{\rm 2loop}(l_0,\boldsymbol{l}) = \frac{e^4 T^2 L^2}{8 \pi ^2 \boldsymbol{l}^2} \ ,\nn\\
&&  \Pi_T^{\rm htl}(l_0,\boldsymbol{l}) = \frac{e^2T^2}{3}\frac{l_0}{4 \boldsymbol{l}^3} \left(2 \vert \boldsymbol{l} \vert l_0-L^2 \ln \left(\frac{l_0+\vert \boldsymbol{l} \vert}{l_0-\vert \boldsymbol{l} \vert}\right)\right) \ ,\nn\\
&& \Pi_T^{\rm pow\cdot corr}(l_0,\boldsymbol{l}) = \frac{e^2}{4 \pi ^2} \left(\frac{l_0^2}{2}+\frac{l_0^4}{6 \boldsymbol{l}^2}
-\frac{2 \boldsymbol{l}^2}{3}
-\frac{l_0^3}{12 \boldsymbol{l}^3}  \left(2 \boldsymbol{l}^2+l_0^2 -\frac{3 \boldsymbol{l}^4}{l_0^2}\right)\ln \left(\frac{l_0+\vert \boldsymbol{l} \vert}{l_0-\vert \boldsymbol{l} \vert}\right)  
\right) \ , \nn\\
&& \Pi_T^{\rm 2loop}(l_0,\boldsymbol{l}) = -\frac{e^4  T^2 }{16 \pi ^2 } \dfrac{l_0}{\vert \boldsymbol{l} \vert} \ln \left(\frac{l_0+\vert \boldsymbol{l} \vert}{l_0-\vert \boldsymbol{l} \vert}\right) \,.
\label{fullPi}
 \ea
 For simplicity, $\Pi_I^{\rm pow\cdot corr}$ above is taken at the value of the renormalization scale $\nu=Te^{-\gamma_E/2-1}\sqrt{\pi}/2$ in the MS scheme. 
This fixes the scale of $e^2=e^2(\nu)$ in $\Pi_I^{\rm htl}$. 

Let us recall the meaning of every term in Eq.~(\ref{exp-HTL}). While the HTL contribution is proportional to $e^2 T^2$, 
the results computed in this manuscript, even if they do not depend on the temperature,  should be viewed as a
a correction of order $m^2/T^2$ to the HTL. Similarly, the power corrections are of order $l^2/T^2$ respect to the HTL, while the two-loop
results are corrections of order $e^2$. These three corrections are of the same order if $m, l \sim eT$, and should be equally
considered. However, if the mass is such that $ e T < m \ll T$, then the mass corrections are dominant at soft scales, $l \sim e T$.

For example, let us take the value of the photon screening mass, defined as $-m_S^2=\Pi_L(l^0=0, \boldsymbol{l}^2=-m_S^2)$.
Calculated from the value of the longitudinal part of the polarization tensor, as given in Eq.~(\ref{exp-HTL}),
 results in 
\be
m_S^2=\frac {e^2 T^2}{3}\left(1-\frac{e^2}{8 \pi^2} - \frac{1}{2 \pi^2} \frac{m^2}{T^2}\right)\, .
\ee
Note that for values $ m^2/T^2 > \pi \alpha$, where $\alpha$ is the
electromagnetic fine structure constant $\sim 1/137$,
the mass effects give   the most important corrections.

Let us finally remind here that while we focused our discussion on thermal plasmas, our results and also the
power corrections of Ref.~\cite{Manuel:2016wqs,Carignano:2017ovz} remain valid in the presence of a chemical potential, or even for
high $\mu$ and $T=0$.
  Our results can also be easily
generalized to QCD, for the mass corrections to the HTL gluon polarization tensor, after taking into account some color factors, and
replacing $e^2$ by $ g^2/2$ for fermions in the fundamental representation, where $g$ is the QCD coupling constant.

Our results might be useful to obtain a better evaluation of different physical observables whenever the fermions in the plasma are not strictly massless, which is a realistic condition for most of the physical scenarios where the HTL resummation techniques have been applied so far.

\section*{Acknowledgements}

We thank J.M~Torres-Rincon for discussions on extensive parts of this project, and S. Carignano and J. Soto for a critical reading of the manuscript.
  We have been supported by the Ministerio de Ciencia  e Innovaci\'on (Spain) under the project  PID2019-110165GB-I00 (MCI/AEI/FEDER, UE),  as well as by the project  2017-SGR-929  (Catalonia).  This work was also supported by the COST Action CA15213 THOR.

\end{document}